\documentclass[aps,prd,eqsecnum,twocolumn,amsfonts,showpacs]{revtex4}
\usepackage{graphics}

\newcommand{\beq}{\begin{equation}}
\newcommand{\eeq}{\end{equation}}
\newcommand{\beqs}{\begin{eqnarray}}
\newcommand{\eeqs}{\end{eqnarray}}
\newcommand{\lsim}{\mathrel{\raisebox{-
.6ex}{$\stackrel{\textstyle<}{\sim}$}}}

\begin{document}

\title{Extended Technicolor Models with Two ETC Groups}

\author{Neil D. Christensen}
\author{Robert Shrock}

\affiliation{C.N. Yang Institute for Theoretical Physics \\
State University of New York, Stony Brook, NY 11794}

\begin{abstract}

We construct extended technicolor (ETC) models that can produce the large
splitting between the masses of the $t$ and $b$ quarks without necessarily
excessive contributions to the $\rho$ parameter or to neutral flavor-changing
processes.  These models make use of two different ETC gauge groups, such that
left- and right-handed components of charge $Q=2/3$ quarks transform under the
same ETC group, while left- and right-handed components of charge $-1/3$ quarks
and charged leptons transform under different ETC groups.  The models thereby
suppress the masses $m_b$ and $m_\tau$ relative to $m_t$, and $m_s$ and $m_\mu$
relative to $m_c$ because the masses of the $Q=-1/3$ quarks and charged leptons
require mixing between the two ETC groups, while the masses of the $Q=2/3$
quarks do not.  A related source of the differences between these mass
splittings is the effect of the two hierarchies of breaking scales of the two
ETC groups.  We analyze a particular model of this type in some detail.
Although we find that this model tends to suppress the masses of the first two
generations of down-type quarks and charged leptons too much, it gives useful
insights into the properties of theories with more than one ETC group. 

\end{abstract}



\maketitle

\vspace{16mm}

\newpage
\pagestyle{plain}
\pagenumbering{arabic}

\section{Introduction}
\label{intro}

It is possible that electroweak symmetry breaks via the formation of a bilinear
condensate of fermions subject to a new, asymptotically free, strong gauge
interaction, generically called technicolor (TC) \cite{tc}.  To communicate
this symmetry breaking to the standard model (technisinglet) fermions, one
embeds technicolor in a larger, extended technicolor (ETC) theory \cite{etc}
(for some reviews, see \cite{etcrev}). To satisfy constraints from
flavor-changing neutral-current (FCNC) processes, the ETC vector bosons that
mediate generation-changing transitions must have large masses. To produce the
hierarchy in the masses of the observed three generations (families) of
fermions, the ETC vector boson masses have a hierarchical spectrum reflecting
the sequential breaking of the ETC gauge symmetry on mass scales ranging from
approximately $10^3$ TeV down to the TeV level. These theories are tightly
constrained by precision electroweak measurements \cite{pdg,lepewwg}.  Modern
technicolor theories are designed so that as the energy scale decreases, the
gauge coupling becomes large but runs very slowly (``walks''); this behavior
enhances masses of standard-model (SM) fermions and pseudo-Nambu-Goldstone
bosons \cite{wtc,chiralpt}.  Models of dynamical electroweak symmetry breaking
are very ambitious in their goals, which include dynamical generation of not
just the $W$ and $Z$ masses but also the entire spectrum of quark and lepton
masses.  It is thus understandable that no fully realistic models of this type
have been developed.  One longstanding challenge for these models has been to
account for the large splitting between the masses of the $t$ and $b$ quarks
without producing excessively large contributions to the parameter
$\rho=m_W^2/(m_Z^2 \cos^2\theta_W)$ or to neutral flavor-changing processes.

In this paper we shall formulate general classes of models that can plausibly
achieve this goal.  These models are based on the general idea suggested in
Ref. \cite{met1}, namely to obtain the splitting of $m_t$ and $m_b$ without
overly large contributions to $\rho$ by using two ETC groups, $G_{ETC}$ and
$G_{ETC}'$ and arranging that the masses of quarks of charge 2/3 involve only
the exchange of gauge bosons belonging to $G_{ETC}$, while the masses of quarks
of charge $-1/3$ arise from diagrams involving gauge bosons of both $G_{ETC}$
and $G_{ETC}'$, requiring mixing between these two sets of gauge bosons.  The
necessity of this mixing suppresses the masses of the down-type quarks relative
to those of up-type quarks.  From among the general classes of ETC models, we
construct and study in some detail one explicit ETC model embodying this idea,
including an analysis of the sequential breaking of the ETC gauge symmetry that
is necessary to produce the hierarchy in the three standard-model fermion
generations.  We also extend the mechanism to leptons so that the mass of the
charged lepton in each generation is suppressed relative to that of the up-type
quark.  A related factor in producing intragenerational mass splittings is the
fact that, for a given generation $j$, the respective breaking scales
$\Lambda_j$ and $\Lambda_j'$ of the $G_{ETC}$ and $G_{ETC}'$ gauge symmetries
may be somewhat different.  We shall focus on a model with one standard-model
family of technifermions (with an additional technineutrino) and also remark on
models in which the left- and right-handed chiral components of the
technifermions transform, respectively as one SU(2)$_L$ doublet and two
SU(2)$_L$ singlets. We also comment on neutrino masses.  Although the model is
rather complicated, we believe that it is useful as an explicit, moderately
ultraviolet-complete realization, of the strategy of using two different ETC
groups to account for the splitting between $m_t$ and $m_b$.

Before proceeding, we briefly review some past efforts to address the problem
of $t$-$b$ mass splitting in ETC theories.  One approach used a one-family
technicolor model and SU(2)$_L$-singlet, charge $-1/3$ vectorlike $b'$ quarks
which mix with the $b$ quark and reduce its mass relative to that of the $t$
quark \cite{at94}.  However, this model and similar ones with $b'$ quarks do
not satisfy the criteria for the diagonality of the hadronic weak neutral
current in terms of mass eigenstates, viz., that all quarks of a given
chirality have the same weak isospin $T$ and $T_3$ \cite{gw}.  Consequently,
such models generically can have problems with excessively large contributions
to hadronic flavor-changing neutral current processes.  A different approach to
the problem of splitting the $t$ and $b$ masses while maintaining acceptably
small corrections to the $\rho$ parameter used a single ETC gauge group with
different ETC representations for the left- and right-handed chiral components
of the down-type quarks and charged leptons \cite{ssvz,jt,ckm,kt}.  However, as
we showed in Refs. \cite{ckm,kt}, this approach also encounters problems with
(i) excessive suppression of down-quark and charged lepton masses, and (ii)
excessively large contributions to flavor-changing neutral current processes,
in particular, $K^0-\bar K^0$ mixing. Related discussions are contained in our
Refs. \cite{dml,qdml}.

This paper is organized as follows.  In Section II we describe the general
structure of one-family models with two ETC gauge groups. In Section III we
construct and study a specific model of this type, including the sequential
breakings of the ETC symmetry groups and the generation of quark and lepton
masses. Section IV contains remarks on some other phenomenological issues such
as flavor-changing neutral current processes, neutrino masses, and the 
minimization of the $S$ parameter. Section V contains our conclusions.

\section{Models with Two ETC Gauge Groups}
\label{generalmodels}

\subsection{Gauge Group}

We take the technicolor group to be SU($N_{TC}$) with the minimal nonabelian
value, $N_{TC}=2$.  There are several reasons for this choice: (i) it reduces
technicolor contributions to the electroweak $S$ parameter describing heavy
fermion loop corrections to the $Z$ boson propagator
\cite{pt,scalc1,scalc2,sred} (a perturbative estimate of which is proportional
to $N_{TC}$ for technifermions in the fundamental representation of
SU($N_{TC}$)); (ii) with eight or nine vectorially coupled Dirac
technifermions, it can plausibly have the desired walking behavior
\cite{wtc,chiralpt}; and (iii) it makes possible a mechanism for obtaining
light neutrino masses \cite{nt,lrs}. The walking behavior can occur naturally
as a result of an approximate infrared-stable fixed point which is larger than,
but close to, a critical value at which the technicolor theory would go over
from a confined phase with spontaneous chiral symmetry breaking to a nonabelian
Coulomb phase \cite{wtc,chiralpt}.

We take the technifermions to transform according to the fundamental
representation of SU(2)$_{TC}$ and focus mainly on models in which they
comprise one standard-model family.  This SU(2)$_{TC}$ arises dynamically from
the sequential breaking of the two ETC groups.  As will be shown below, the
last stage of this sequence, which yields the technicolor group, entails the
breaking of a direct product group ${\rm SU}(2)_{ETC} \times {\rm
SU}(2)'_{ETC}$ to its diagonal subgroup.  To embed this direct product group
in the two respective larger ETC groups, taken to be $G_{ETC} = {\rm
SU}(N_{ETC})$ and $G_{ETC}' = {\rm SU}(N_{ETC}')$, one gauges the generational
indices and combines them with the two sets of SU(2) group indices, leading to
the relation
\beq
N_{ETC}=N_{ETC}' = N_{gen.}+N_{TC} \ , 
\label{nrel}
\eeq
where $N_{gen.}=3$ is the number of SM fermion generations.  With $N_{TC}=2$,
this then yields 
\beq
G_{ETC} = {\rm SU}(5)_{ETC} \ , \quad G_{ETC}' = {\rm SU}(5)'_{ETC} \ . 
\label{getcgetcp}
\eeq
Thus, in this class of models the meaning of a standard-model generation is
somewhat different from the meaning in a conventional ETC model with only a
single ETC group; here, for down-type quarks and charged leptons, there are
really two kinds of generations for the two chiralities, corresponding to
different ETC gauge groups.  We shall also use additional strongly coupled
gauge interactions to produce the desired sequential ETC symmetry-breaking
pattern. These include two SU(2) hypercolor gauge interactions, each
corresponding to a different ETC group, denoted SU(2)$_{HC}$ and SU(2)$'_{HC}$
\cite{hc}.  The ETC symmetry breaking occurs in sequential stages: (i)
SU(5)$_{ETC}$ breaks to SU(4)$_{ETC}$ at a scale denoted $\Lambda_1$ and
SU(5)$'_{ETC}$ breaks to SU(4)$'_{ETC}$ at a scale $\Lambda_1'$, where the
subscript 1 is assigned because it is at this stage that the first-generation
quarks and leptons split off from the remaining four components in fundamental
representations of SU(5)$_{ETC}$ and SU(5)$'_{ETC}$; (ii) SU(4)$_{ETC}$ and
SU(4)$'_{ETC}$ break to SU(3)$_{ETC}$ and SU(3)$'_{ETC}$, respectively, at the
lower scales scales $\Lambda_2$ and $\Lambda_2'$ where the second-generation
fermions split off from the remaining three components of the above
representations; and (iii) SU(3)$_{ETC}$ and SU(3)$'_{ETC}$ break to
SU(2)$_{ETC}$ and SU(2)$'_{ETC}$, respectively, at the lower scales 
$\Lambda_3$ and $\Lambda_3'$, where third-generation fermions split off from
the remaining two components of the above representations. 

A basic requirement for these models is that there must be a mechanism for
communicating between SU(5)$_{ETC}$ and SU(5)$'_{ETC}$ in order to produce an
SU(2)$_{TC}$ group that leads to a technifermion condensate.  This mechanism
must involve all of the five indices of both the SU(5)$_{ETC}$ and
SU(5)$'_{ETC}$ in order to give masses to all three generations of fermions.
{\it A priori}, one could consider several possible ways of producing this
communication, thereby defining corresponding classes of models.  In the models
that we shall study here, this communication is achieved by the use of a
strongly coupled gauge interaction, which we shall call metacolor (MC), which
mediates between the ETC and ETC$'$ groups.  In the explicit models considered
here, we take the metacolor gauge group to be SU(2)$_{MC}$.  The communication
is effected by metacolor-induced condensates of a set of SM-singlet fermions
that transform as nonsinglets under SU(5)$_{ETC}$ and SU(2)$_{MC}$ with another
set that transform as nonsinglets under SU(5)$'_{ETC}$ and SU(2)$_{MC}$.  We
denote this class of ETC models as EMC for ``ETC with MC'' \cite{efc}.

In an EMC type of ETC model, starting with a low-energy effective field theory
involving descendents of the original ETC groups, which we will denote as
$H_{ETC} \times H_{ETC}'$, with $H_{ETC} \subset {\rm SU}(5)_{ETC}$ and
$H_{ETC}' \subset {\rm SU}(5)_{ETC}'$, the metacolor interaction produces
the breaking $H_{ETC} \times H_{ETC}' \to {\rm SU}(2)_{TC}$.  We denote the
scale of this breaking as $\Lambda_{MC}$, which is also the scale where the
metacolor gauge interaction becomes strong.  This scale should be smaller than
each of the lowest generational symmetry-breaking scales in the two respective
ETC sector, which are involved in the dynamical production of third-generation
fermion masses, $\Lambda_3$ and $\Lambda_3'$, since if this were not the case,
i.e., if there were a single ETC group operative at a scale $\ge \Lambda_3$,
then the basic mechanism considered here for splitting the $t$ and $b$ masses
would not be operative.  Since $\Lambda_3 < \Lambda_j$ for $j=1,2$ and
$\Lambda_3' < \Lambda_j'$, $j=1,2$, it follows that
\beq
\Lambda_{MC} < {\rm min}(\Lambda_3, \ \Lambda_3') \ . 
\label{lammcineq}
\eeq
In turn, this implies that in the above discussion, 
$H_{ETC} = {\rm SU}(2)_{ETC}$ and $H_{ETC}' = {\rm SU}(2)_{ETC}'$. 

Thus, the respective full gauge group of the fundamental theory is 
\beqs
& & G =  {\rm SU}(5)_{ETC} \times {\rm SU}(5)'_{ETC} \cr\cr
& \times &  
{\rm SU}(2)_{HC} \times {\rm SU}(2)'_{HC}
 \times {\rm SU}(2)_{MC} \times G_{SM}
\label{g}
\eeqs
where $G_{SM} = {\rm SU}(3)_c \times {\rm SU}(2)_L \times {\rm U}(1)_Y$ is the
standard-model gauge group with $N_c=3$.

\subsection{Standard-Model Fermion Content}

We next discuss the choices of standard-model fermion representations under the
ETC and ETC$'$ gauge groups.  We assign the left-handed quark and techniquark
SU(2)$_L$ doublets to transform as a fundamental representation of
SU(5)$_{ETC}$ and a singlet under SU(5)$'_{ETC}$.  (Our choice of a fundamental
rather than conjugate fundamental representation of SU(5)$_{ETC}$ is a
convention.) Since we want the up-type and down-type quark masses in each of
the second and third generations to be unsuppressed and suppressed,
respectively, it follows that the right-handed components of the up- and
down-type quarks and techniquarks should be assigned to the (5,1) and (1,5)
representations of ${\rm SU}(5)_{ETC} \times {\rm SU}(5)'_{ETC}$, respectively.
The fermions that are nonsinglets under $G_{SM}$ are singlets under the
hypercolor and the metacolor group.  This then determines the
quark sector of our model, which is displayed below, where the numbers indicate
the representations under the nonabelian factor groups in $G$, and the
subscript gives the weak hypercharge $Y$:
\beqs 
& & Q_L: \ (5,1; \{1\};3,2)_{1/3,L} \cr\cr
& & u_R: \ (5,1; \{1\};3,1)_{4/3,R} \cr\cr
& & d_R: \ (1,5; \{1\};3,1)_{-2/3,R}
\label{1fam_quarks} 
\eeqs
where $\{1\}$ means singlet under ${\rm SU}(2)_{HC} \times {\rm SU}(2)'_{HC}
\times {\rm SU}(2)_{MC}$.  Here and in the rest of the paper we use a compact
notation in which, for example,
\beqs
u_R \equiv u^{aj}_R & \equiv & (u^{a1},u^{a2},u^{a3},u^{a4},u^{a5})_R \cr\cr
                    & \equiv & (u^a,c^a,t^a,U^{a4},U^{a5})_R
\label{uexample}
\eeqs
where $a$ and $j$ are color and SU(5)$_{ETC}$ indices, and so forth for the
other fields.  One could also consider a model in which $d_R$ transforms as
$(1,\bar 5;1,1,1;3,1)_{-2/3,R}$ rather than $(1,5;1,1,1;3,1)_{-2/3,R}$. This
would entail further reduction of the down-quark masses; we shall focus here on
the choice in eq. (\ref{1fam_quarks}).  (In passing, we note that the
assignment of $d_R$ to $(\bar 5,1;1,1,1;3,1)_{-2/3,R}$ would produce a model
similar to the one that we studied in Refs. \cite{ckm,kt}; while this would
succeed in intragenerational reduction of down-quark masses relative to
up-quark masses, it would yield overly large ETC contributions to
flavor-changing neutral current processes.)

In order to suppress the charged lepton mass relative to the up-quark mass in
each generation, we shall make the left- and right-handed components of the
charged leptons transform according to different ETC groups. Given our
restriction to fundamental and conjugate fundamental representations, 
we shall consider two different cases, which we label as $L1$ and $L2$:
\beqs
{\rm case} \ L1: & & L_L: \ (1,5;\{1\};1,2)_{-1,L} \cr\cr
& & e_R: \ (5,1;\{1\};1,1)_{-2,R} 
\label{1fam_leptons1} 
\eeqs
and 
\beqs
{\rm case} \ L2: & & L_L: \ (1,\bar 5;\{1\};1,2)_{-1,L} \cr\cr
& & e_R: \ (\bar 5,1;\{1\};1,1)_{-2,R}  \ . 
\label{1fam_leptons2} 
\eeqs
One can then characterize a given model as being of type $L1$ or $L2$.

\subsection{Anomaly Constraints on SM-Singlet Fermion Content} 

A requirement in the construction of these models is the absence of any local
gauge anomalies and, for the SU(2) groups, also the absence of any global
$\pi_4$ anomalies.  For the model with lepton assignments of type L1,
SM-nonsinglet fermions and technifermions contribute the following terms to
these gauge anomalies: (i) for the cubic SU(5)$_{ETC}$ anomaly (written for
right-handed chiral components), $A(Q^c_R)=-2N_c=-6$, $A(u_R)=N_c=3$, and
$A(e_R)=1$, for a total of $-2$; (ii) for the cubic SU(5)$'_{ETC}$ anomaly
(again for right-handed chiral components): $A(L^c)_R=-2$ and $A(d_R)=3$, for a
total of 1. Here and below, we always write standard-model singlet fields as
right-handed.  We cancel the above cubic anomalies for SU(5)$_{ETC}$ and
SU(5)$'_{ETC}$ with SM-singlet, SU(5)$_{ETC}$-nonsinglet fermions $f_R$, which
contribute the amounts
\beq 
{\rm case} \ L1: \ 
\sum_{f_R} A(f_R) =  2 \ , \quad \sum_{\tilde f_R} A'(\tilde f_R) =  -1 \ . 
\label{arsum_1fam_L1}
\eeq
For the model with lepton assignment L2, by similar reasoning, we
have the constraint
\beq 
{\rm case} \ L2: \ 
\sum_{f_R} A(f_R)=4 \ , \quad \sum_{\tilde f_R} A'(\tilde f_R)=-5 \ .
\label{arsum_1fam_L2}
\eeq
One can check that these models also have zero SU(5)$^2$U(1)$_Y$ and 
SU(5)$'^2$U(1)$_Y$ gauge anomalies. 

Before proceeding, we remark on the corresponding constraints for gauge anomaly
cancellation in the effective field theories that result from the sequential
breaking of the ${\rm SU}(5)_{ETC} \times {\rm SU}(5)'_{ETC}$ symmetry to ${\rm
SU}(N)_{ETC} \times {\rm SU}(N')'_{ETC}$, where $N$ and $N'$ decrease through
the values 4 and 3.  The origin of the conditions
(\ref{arsum_1fam_L1})-(\ref{arsum_1fam_L2}) is the contributions of the
standard-model fermion multiplets to the respective ETC and ETC$'$ anomalies,
and since these SM-fermion ETC multiplets are fundamental representations for
each of the descendent ETC and ETC$'$ groups, their anomaly contributions
remain the same.  It follows that for each of the low-energy effective field
theories invariant under the various ${\rm SU}(N)_{ETC} \times {\rm
SU}(N')'_{ETC}$ gauge groups with $N$ and $N'$ taking on values 4 and 3, the
generalizations of conditions (\ref{arsum_1fam_L1})-(\ref{arsum_1fam_L2}) hold,
where now $A$ and $A'$ denote the respective contributions to the ${\rm
SU}(N)_{ETC}$ and ${\rm SU}(N')'_{ETC}$ anomalies from the massless SM-singlet
fermions $f_R$ and $\tilde f_R$ which are nonsinglets under these two groups.
(That is, in the respective sums $\sum_{f_R}$ and $\sum_{\tilde f_R}$, one has
removed fermions that have gained dynamical masses due to the formation of
condensates at higher scales.)  The SM-singlet sectors of the descendant
low-energy effective field theories resulting from the sequential symmetry
breaking of the ETC and ETC' symmetries satisfy these anomaly constraints, as
can be checked explicitly.

\section{A Model of Type $L1$}
\label{model1}

\subsection{Standard-Model Singlet Fermion Content}

Here we construct and analyze a model of type $L1$. 
We must first choose an SM-singlet fermion sector that
satisfies the anomaly constraints of eq. (\ref{arsum_1fam_L1}).  Here we use
one relatively simple solution to these constraints for the SM-singlet
fermions:
\beqs
& & \psi^{ij}_R \ : \            (10,1;1,1,1;1,1)_{0,R} \cr\cr 
& & {\cal N}^i_R \: \             (5,1;1,1,1;1,1)_{0,R} \cr\cr 
& & \zeta^{ij,\alpha}_R \  :     (10,1;2,1,1;1,1)_{0,R} \cr\cr
& & \omega^\alpha_{p,R} \ :    2(1,1;2,1,1;1,1)_{0,R} \cr\cr
& & \chi_{i,\lambda,R} \ : \   (\bar 5,1;1,1,2;1,1)_{0,R} 
\label{f_EMMEL1}
\eeqs
and
\beqs
& & \tilde \psi_{i'j',R} \ : \ (1,\overline{10};1,1,1;1,1)_{0,R} \cr\cr
& & \tilde \zeta_{i'j',\alpha',R} \ : \ (1,\overline{10};1,2,1;1,1)_{0,R}\cr\cr
& & \tilde \omega_{\alpha',p,R} \ : \quad 2(1,1;1,2,1;1,1)_{0,R} \cr\cr
& & \tilde \chi^{i',\lambda}_R \ : \ (1,5;1,1,2;1,1)_{0,R} \ , 
\label{ftilde_EMMEL1}
\eeqs
where $1 \le i,j \le 5$ are SU(5)$_{ETC}$ indices, $1 \le i',j' \le 5$ are
SU(5)$'_{ETC}$ indices, $\alpha=1,2$ are SU(2)$_{HC}$ indices, $\alpha'=1,2$
are SU(2)$'_{HC}$ indices, $\lambda=1,2$ are SU(2)$_{MC}$ indices, and $p=1,2$
refer to the two copies of the fields $\omega^\alpha_{p,R}$ and $\tilde
\omega^{\alpha'}_{p,R}$. (It is necessary to have an even number of copies of
the $\omega^\alpha_{p,R}$ and $\tilde \omega^\alpha_{p,R}$ fields in order to
avoid global $\pi_4$ anomalies in the SU(2)$_{HC}$ and SU(2)$'_{HC}$ gauge
sectors.)  In the above equations, the tilde denotes SM-singlet,
SU(5)$_{ETC}$-singlet, SU(5)$'_{ETC}$-nonsinglet fermions, and we use the fact
that the representations of SU(2) are (pseudo)real.  At certain points below,
we shall denote technicolor indices by $t$ to distinguish them from
generational indices $j,k \in \{1,2,3\}$.

With this content of massless fermions, the SU(5)$_{ETC}$, SU(5)$'_{ETC}$,
SU(2)$_{HC}$, SU(2)$'_{HC}$, and SU(2)$_{MC}$ interactions are all
asymptotically free.  The leading coefficients of the various beta functions
are given by \cite{beta} 
\beq
b_0^{(SU(5)_{ETC})}=11 \ , \quad b_0^{(SU(5)'_{ETC})}=13 \ ,
\label{b0etc_model1}
\eeq
\beq
b_0^{(SU(2)_{HC})}=b_0^{(SU(2)'_{HC})}=\frac{10}{3} \ , 
\label{b0hc_model1}
\eeq
and
\beq
b_0^{(SU(2)_{MC})}=4 \ . 
\label{b0mc_model1}
\eeq
The beta functions for the other two nonabelian groups in $G$, SU(3)$_c$,
SU(2)$_L$, are also asymptotically free, with leading coefficients
$b_0^{(SU(3)_c)}=13/3$ and $b_0^{(SU(2)_L)}=2/3$.  These values of beta
function coefficients for standard-model gauge groups also hold for all of the
other 1-family models considered here, since they all have the same
content of SM-nonsinglet fermions.

\subsection{General Structure of ETC Gauge Symmetry Breaking} 

The ETC and ETC$'$ gauge symmetries are chiral, so that when they become
strong, sequential breaking of each occurs naturally \cite{tumb}.  This
breaking also involves additional strongly coupled gauge interactions.  The
breakings of the SU(5)$_{ETC}$ to SU(2)$_{ETC}$ and of SU(5)$'_{ETC}$ to
SU(2)$'_{ETC}$ are driven by the condensation of SM-singlet fermions.  The
SM-singlet fermions that condense at a given scale acquire dynamical masses of
order this scale and hence decouple from the effective theory at lower
energies.

We identify plausible preferred condensation channels using a generalized
most-attractive-channel (GMAC) approach that takes account of one or more
strong gauge interactions at each breaking scale, as well as the energy cost
involved in producing gauge boson masses when gauge symmetries are broken. An
approximate measure of the attractiveness of a channel $R_1 \times R_2 \to
R_{cond.}$ is $\Delta C_2 = C_2(R_1)+C_2(R_2)-C_2(R_{cond.})$ \ \cite{gap},
where $R_j$ denotes the representation under a relevant gauge interaction and
$C_2(R)$ is the quadratic Casimir invariant \cite{casimir}.

\subsection{SU(5)$_{ETC} \to$ SU(4)$_{ETC}$ and SU(5)$'_{ETC} \to$ 
SU(4)$'_{ETC}$ Breaking} 

As the energy scale decreases from high values, the SU(5)$_{ETC}$ coupling
increases, as governed by the coefficient $b_0^{(SU(5)_{ETC})}$ in
eq. (\ref{b0etc_model1}).  We envision that at an energy scale $\Lambda_1$ of
order $10^3$ TeV, $\alpha_{_{ETC}}$ becomes sufficiently large to cause a
condensate in the channel
\beqs
& & (10,1;1,1,1;1,1)_{0,R} \times (10,1;1,1,1;1,1)_{0,R}
\to \cr\cr
& & \to (\bar 5,1;1,1,1;1,1)_0
\label{1010channel_model1}
\eeqs
with $\Delta C_2 = 24/5$, breaking ${\rm SU}(5)_{ETC}$ to ${\rm SU}(4)_{ETC}$.
This is a most attractive channel; i.e., there is no other channel with a
higher value of $\Delta C_2$ \cite{other1}. With no loss of generality, we take
the breaking direction in SU(5)$_{ETC}$ as $i=1$, thereby splitting off the
first-generation up quarks from the remaining components of the corresponding
SU(5) fields with indices $i \in \{2,3,4,5\}$.  With respect to the unbroken
${\rm SU}(4)_{ETC}$, we have the decomposition $10 = 4 + 6$. (Note that $6
\approx \bar 6$ in SU(4).)  We denote the fundamental representation of
SU(4)$_{ETC}$, $(4,1;1,1,1;1,1)_{0,R}$, as $\alpha^{1i}_R \equiv \psi^{1i}_R$
for $2 \le i \le 5$, and the antisymmetric tensor representation
$(1,6;1,1,1;1,1)_{0,R}$ as $\xi^{ij}_R \equiv \psi^{ij}_R$ for $2 \le i,j \le
5$. The associated condensate is
\beq 
\langle \epsilon_{1 i j k \ell} \xi^{ij \ T}_R C \xi^{k \ell}_R \rangle
= 8\langle 
\xi^{23 \ T}_R C \xi^{45}_R - 
\xi^{24 \ T}_R C \xi^{35}_R +
\xi^{25 \ T}_R C \xi^{34}_R \rangle \ . 
\label{xixicondensate_model1}
\eeq
where here and below, summations over repeated indices are understood.  Linear
combinations of the six $\xi^{ij}_R$ fields involved in this condensate pick up
masses of order $\Lambda_1$. 

Similarly, at a comparable scale $\Lambda_1' \simeq 10^3$ TeV, we envision that
$\tilde \alpha_{_{ETC}}$ becomes sufficiently strong to produce condensation in
the channel
\beqs
& & (1,\overline{10};1,1,1;1,1)_{0,R} \times 
    (1,\overline{10},;1,1,1;1,1)_{0,R} \to \cr\cr
& & \to (1,5;1,1,1;1,1)_0
\label{10b-10bchannel_model1}
\eeqs
breaking SU(5)$'_{ETC}$ to SU(4)$'_{ETC}$. Again, this is a most attractive
channel, with $\Delta C_2=24/5$.  With no loss of generality, we take the
breaking direction in SU(5)$'_{ETC}$ as $i'=1$; this entails the separation of
the first generation of down-type quarks and charged leptons from the
components of the corresponding SU(5)$'_{ETC}$ fields with indices lying in the
set $\{2,3,4,5\}$.  In analogy with our notation for SU(4)$_{ETC}$, we denote
the conjugate fundamental representation $(1,\bar 4;1,1,1;1,1)_{0,R}$ and
antisymmetric conjugate tensor representation $(1,\bar 6;1,1,1;1,1)_{0,R}$ of
SU(4)$'_{ETC}$ as $\tilde \alpha_{1i,R} \equiv \tilde \psi_{1i,R}$ for $2 \le i
\le 5$ and $\tilde \xi_{ij,R} \equiv \tilde \psi_{ij,R}$ for $2 \le i,j \le 5$.
The associated SU(5)$'_{ETC}$-breaking, SU(4)$'_{ETC}$-invariant condensate is
\beqs
\langle \epsilon^{1i'j'k' \ell'}
\tilde \xi^T_{i'j',R} C \tilde \xi_{k' \ell',R} \rangle & = & 8\langle 
\tilde \xi^T_{23,R} C \tilde \xi_{45,R} - 
\tilde \xi^T_{24,R} C \tilde \xi_{35,R} \cr\cr  
& + & \tilde \xi^T_{25,R} C \tilde \xi_{34,R} \rangle \ . 
\label{xixitildecondensate} 
\eeqs
The six $\tilde \xi_{i'j',R}$ fields involved in this condensate pick up masses
of order $\Lambda_1'$. (Again, the actual mass eigenstates are linear
combinations of these fields; henceforth, we shall often suppress this when it
is not important for the discussion.)  As was true in our previous studies of
ETC models, at lower energy scales, different patterns of ETC breaking can
occur, depending on the relative strengths of the ETC and HC gauge couplings.
We shall focus on one pattern here.

\subsection{ETC Symmetry Breaking at Lower Mass Scales}

In the energy interval just below the lower of the two scales $\Lambda_1$ and
$\Lambda_1'$, the effective theory is invariant under the gauge group
\beqs
& & {\rm SU}(4)_{ETC} \times {\rm SU}(4)'_{ETC} \times {\rm SU}(2)_{HC} \times
{\rm SU}(2)'_{HC} \times \cr\cr
& & \times {\rm SU}(2)_{MC} \times G_{SM} \ . 
\label{g44}
\eeqs
Since the SU(4)$_{ETC}$ and SU(4)$'_{ETC}$ gauge interactions are
asymptotically free, the corresponding couplings $\alpha_{_{ETC}}$ and $\tilde
\alpha_{_{ETC}}$ continue to increase as the energy scale decreases.  As the
energy scale descends through the value $\Lambda_2 \simeq 50$ TeV, the
SU(4)$_{ETC}$ and SU(2)$_{HC}$ couplings become sufficiently strong to lead
together to condensation in the channel
\beqs
& & (4,1;2,1,1;1,1)_{0,R} \times (6,1;2,1,1;1,1)_{0,R} \to 
\cr\cr
& & (\bar 4,1;1,1,1;1,1)_0  \ . 
\label{4x6channel} 
\eeqs
This condensation channel preserves SU(2)$_{HC}$ and breaks SU(4)$_{ETC}$ to
SU(3)$_{ETC}$.  The Casimir operators measuring the attractiveness of this
channel are $\Delta C_2 = 5/2$ for SU(4)$_{ETC}$ and $\Delta C_2 = 3/2$ for 
SU(2)$_{HC}$.  The associated condensate is
\beqs 
& & \langle \epsilon_{\alpha\beta}\epsilon_{12jk \ell}\zeta^{1j,\alpha\ T}_R 
C \zeta^{k \ell,\beta}_R \rangle = 2\langle
\epsilon_{\alpha\beta}( \zeta^{13,\alpha \ T}_R C \zeta^{45,\beta}_R \cr\cr
& & -
\zeta^{14,\alpha \ T}_R C \zeta^{35,\beta}_R +
\zeta^{15,\alpha \ T}_R C \zeta^{34,\beta}_R ) \rangle \ , 
\label{4x6zetacondensate} 
\eeqs
and the twelve $\zeta^{ij,\alpha}_R$ fields in this condensate gain
masses of order $\Lambda_2$.  

Analogously, as the energy scale decreases through the value 
$\Lambda_2' \simeq 15$ TeV, the SU(4)$'_{ETC}$ and
SU(2)$'_{HC}$ couplings become sufficiently strong to lead together to
condensation in the channel
\beqs
& & (1,\bar 4;1,2,1;1,1)_{0,R} \times (1,\bar 6;1,2,1;1,1)_{0,R} \to 
\cr\cr
& & \to (1,4;1,1,1;1,1)_0
\label{4bx6bchannel} 
\eeqs
breaking SU(4)$'_{ETC}$ to SU(3)$'_{ETC}$.  The quadratic Casimir invariants
for this channel are $\Delta C_2 = 5/2$ for SU(4)$'_{ETC}$ and $\Delta C_2 =
3/2$ for SU(2)$'_{HC}$.  The condensate is
\beqs 
& & \langle \epsilon^{\alpha'\beta'} \, \epsilon^{12j'k' \ell'} \, 
\tilde \zeta^T_{1j',\alpha',R} C \tilde \zeta_{k' \ell',\beta',R} \rangle =
\cr\cr & = & 2\langle \epsilon_{\alpha'\beta'}( 
\tilde \zeta^T_{13,\alpha',R} C \tilde \zeta_{45,\beta',R} \cr\cr 
& & 
- \tilde \zeta^T_{14,\alpha',R} C \tilde \zeta_{35,\beta',R} +
\tilde \zeta^T_{15,\alpha',R} C \tilde \zeta_{34,\beta',R} ) \rangle \ , \cr\cr
& & 
\label{4bx6bzetatildecondensate} 
\eeqs
and the twelve $\tilde \zeta_{i'j',\alpha',R}$ fields in this condensate gain
masses $\sim \Lambda_2'$.

The effective theory just below $\Lambda_2'$ is invariant under the gauge
group
\beqs
& & {\rm SU}(3)_{ETC} \times {\rm SU}(3)'_{ETC} \times {\rm SU}(2)_{HC} \times
{\rm SU}(2)'_{HC} \times \cr\cr
& & \times {\rm SU}(2)_{MC} \times G_{SM} \ . 
\label{g33}
\eeqs
Since the SU(3)$_{ETC}$, SU(3)$'_{ETC}$, SU(2)$_{HC}$, SU(2)$'_{HC}$, and
SU(2)$_{MC}$ interactions are asymptotically free, their couplings continue to
increase as the energy scale decreases.  At the scale $\Lambda_3$ of a few TeV,
the SU(3)$_{ETC}$ and SU(2)$_{HC}$ interactions trigger condensation in the
channel
\beqs
& & (3,1;2,1,1;1,1)_{0,R} \times (3,1;2,1,1;1,1)_{0,R} \to \cr\cr
& & \to (\bar 3,1;1,1,1;1,1)_0 \ , 
\label{33to3bchannel} 
\eeqs
where the numbers indicate the representations under the group (\ref{g33}).
This condensation is invariant under SU(2)$_{HC}$ and breaks SU(3)$_{ETC}$ to
SU(2)$_{ETC}$.  Its attractiveness is given by the Casimir invariants
$\Delta C_2 = 4/3$ for SU(3)$_{ETC}$ and $\Delta C_2 = 3/2$ for SU(2)$_{HC}$.
Without loss of generality, we may use the original SU(3)$_{ETC}$ gauge
symmetry to orient the condensate so that it takes the form
\beq 
\langle \epsilon_{123jk} \epsilon_{\alpha\beta} \zeta^{2j,\alpha \ T}_R
C \zeta^{2k,\beta}_R \rangle \ = \ 2\langle \epsilon_{\alpha\beta}
 \zeta^{24,\alpha}_R C  \zeta^{25,\beta}_R \rangle \ . 
\label{33to3bcondensate} 
\eeq

Similarly, at a scale $\Lambda_3'$, the SU(3)$'_{ETC}$ and SU(2)$'_{HC}$
are envisioned to lead together to a condensation in the channel
\beqs
& & (1,\bar 3;1,2,1;1,1)_{0,R} \times (1,\bar 3;1,2,1;1,1)_{0,R} \to \cr\cr
& & \to (1,3;1,1,1;1,1)_0 \ , 
\label{3b3bto3channel} 
\eeqs
breaking SU(3)$'_{ETC}$ to SU(2)$'_{ETC}$. The condensate is 
\beq 
\langle \epsilon^{123j'k'} \epsilon^{\alpha'\beta'} 
\tilde \zeta^T_{2j',\alpha',R}
C \tilde \zeta_{2k',\beta',R} \rangle \ = \ 2\langle \epsilon^{\alpha'\beta'}
\tilde \zeta^T_{24,\alpha',R} C \tilde \zeta_{25,\beta',R} \rangle \ . 
\label{3b3bto3condensate} 
\eeq

We next discuss several additional condensations driven by the SU(2)$_{HC}$ and
SU(2)$'_{HC}$ interactions.  In the low-energy effective field theory just
below min($\Lambda_3,\Lambda_3'$), the massless SM-singlet, ETC-nonsinglet
fermions consist of $\zeta^{ij,\alpha}_R$ and $\tilde\zeta_{i'j',\alpha',R}$
with $ij=12, 23$ and $i'j'=12, 23$, together with $\omega^\alpha_{p,R}$
and $\tilde \omega_{\alpha',p,R}$ with $p=1,2$.  In this energy interval, the
SU(2)$_{ETC}$, SU(2)$'_{ETC}$, and SU(2)$_{HC}$ gauge couplings continue to
grow.  The hypercolor interaction naturally produces (HC-singlet) condensates
of the various remaining HC-doublet fermions.  In each case, $\Delta C_2 =3/2$
for the hypercolor interaction.  Since the condensates (\ref{33to3bcondensate})
and (\ref{3b3bto3condensate}) were formed via a combination of the attractive
SU(2)$_{HC}$ and SU(2)$'_{HC}$ interaction with, respectively, the
SU(3)$_{ETC}$ and SU(3)$'_{ETC}$ interactions, while the present condensates
are formed only by the SU(2)$_{HC}$ or SU(2)$'_{HC}$ interaction, and have the
same value of $\Delta C_2$ for the HC and HC$'$ groups, it follows that the
scales at which they form, denoted $\Lambda_s$ and $\Lambda_s'$ (where $s$
denotes SU(2)$_{ETC}$-singlet and SU(2)$'_{ETC}$-singlet) satisfy (i)
$\Lambda_s \le \Lambda_3$ and $\Lambda_s' \le \Lambda_3'$.  There are twelve
condensates of this type,
\beq 
\langle \epsilon_{\alpha\beta} \zeta^{12,\alpha \ T}_R C
\zeta^{23,\beta}_R \rangle 
\label{z12z23condensate} 
\eeq
\beq 
\langle \epsilon_{\alpha\beta} \zeta^{12,\alpha \ T}_R C
\omega^\beta_{p,R} \rangle
\label{z12omegacondensate} 
\eeq
\beq 
\langle \epsilon_{\alpha\beta} \zeta^{23,\alpha \ T}_R C
\omega^\beta_{p,R} \rangle
\label{z23omegacondensate} 
\eeq
\beq 
\langle \epsilon_{\alpha\beta} \omega^{\alpha \ T}_{1,R} C
\omega^\beta_{2,R} \rangle 
\label{omega_selfcondensate} 
\eeq
\beq 
\langle \epsilon^{\alpha'\beta'} \tilde \zeta^T_{12,\alpha',R}
C \tilde \zeta_{23,\beta',R} \rangle 
\label{z12z23tildecondensate} 
\eeq
\beq 
\langle \epsilon^{\alpha'\beta'} 
\tilde \zeta^T_{12,\alpha',R} C
\tilde \omega_{\beta',p,R} \rangle
\label{z12omegatildecondensate} 
\eeq
\beq 
\langle \epsilon^{\alpha'\beta'} 
\tilde \zeta^T_{23,\alpha',R} C
\tilde \omega_{\beta',p,R} \rangle
\label{z23omegatildecondensate} 
\eeq
\beq 
\langle \epsilon^{\alpha'\beta'} 
\tilde \omega^T_{\alpha',1,R} C
\tilde \omega_{\beta',2,R} \rangle 
\label{omega_tildeselfcondensate} 
\eeq
where $p=1,2$. Here we shall take $\Lambda_s \simeq \Lambda_s' \simeq
\Lambda_3$.

In the energy interval just below ${\min}(\Lambda_3, \Lambda_3')$, the
theory is invariant under the gauge group
\beqs
& & {\rm SU}(2)_{ETC} \times {\rm SU}(2)'_{ETC} \times {\rm SU}(2)_{HC} \times
{\rm SU}(2)'_{HC} \times \cr\cr
& & \times {\rm SU}(2)_{MC} \times G_{SM} \ . 
\label{g22}
\eeqs
With the fermions that have gained dynamical masses at scales $\ge \Lambda_s,
\Lambda_s'$ integrated out, the resultant low-energy effective field theory
contains 14 chiral fermions transforming as doublets under SU(2)$_{ETC}$,
namely
\beq
Q_L^{aj}, \ \  u_R^{aj}, \ \ e_R^j, \ \ \alpha^{1j}_R, \ \ {\cal N}^j_R, \ 
\chi_{j,\lambda,R} 
\label{su2doublets}
\eeq
and eight chiral fermions transforming as chiral doublets under
SU(2)$'_{ETC}$, namely
\beq
L_L^{j'}, \ \ d_R^{aj'}, \ \ \tilde \alpha_{1j',R}, \ \ 
\tilde \chi^{j',\lambda}_R
\label{su2primedoublets}
\eeq
with $j=4,5$, $j'=4,5$, and $\lambda=1,2$. 

As the energy scale decreases through the value $\Lambda_{MC}$, the metacolor
interaction gets sufficiently strong to lead to the condensation of the
metacolor-nonsinglet fermions.  We assume that $\Lambda_{MC}$ is of order a few
TeV. Applying a GMAC argument, we infer that the favored condensation channel,
as regards metacololor, is $2 \times 2 \to 1$, with condensate 
\beq 
\langle \sum_{\lambda=1}^2 \sum_{j,k'=1}^5 \chi^T_{j,\lambda,R} C \tilde
\chi^{k',\lambda}_R \rangle
\label{mccondensate}
\eeq
where $\lambda$ are MC indices, and $j$ and $k'$ are SU(5)$_{ETC}$ and
SU(5)$'_{ETC}$ indices, respectively.  Although these latter two groups,
SU(5)$_{ETC}$ and SU(5)$'_{ETC}$, are no longer invariance groups of the
effective theory at this scale $\Lambda_{MC}$, the range of the indices in the
condensate (\ref{mccondensate}) is still $1 \le j,k' \le 5$ since the
$\chi_{j,\lambda,R}$ and $\tilde \chi^{k',\lambda}_R$ are still massless
fermions at this scale, and this condensate can be bound solely by the
SU(2)$_{MC}$ interaction.  For the components $j,k' \in \{4,5\}$, a vacuum
alignment argument implies that the condensate (\ref{mccondensate}) is of the
form ${\rm const.} \times \delta_j^{k'}$ so that it breaks ${\rm SU}(2)_{ETC}
\times {\rm SU}(2)'_{ETC}$ to the diagonal subgroup ${\rm SU}(2)_d \equiv {\rm
SU}(2)_{TC}$. The components of the $\chi$ and $\tilde \chi$ fermions involved
in this condensate thus gain dynamical masses of order $\Lambda_{MC}$.  The
three normalized diagonal linear combinations of the gauge bosons of ${\rm
SU}(2)_{ETC} \times {\rm SU}(2)'_{ETC}$ are the massless gauge bosons of
SU(2)$_{TC}$ and the three orthogonal linear combinations gain masses of order
$\Lambda_{MC}$.  Thus, the communication between the SU(5)$_{ETC}$ and
SU(5)$'_{ETC}$ groups takes place at the scale $\Lambda_{MC}$.  This
communication, via the condensate, (\ref{mccondensate}) achieves two important
goals: (i) connecting fermions with SU(5)$_{ETC}$ generation indices $j=1,2,3$
and fermions with SU(5)$'_{ETC}$ generation $k'=1,2,3$ indices; and (ii)
producing the exact SU(2)$_{TC}$ group which will break electroweak
interactions at a lower energy scale. 

The effective field theory just below $\Lambda_{MC}$ is thus invariant under
the gauge group
\beq
{\rm SU}(2)_{TC} \times {\rm SU}(2)_{HC} \times {\rm SU}(2)'_{HC} 
\times {\rm SU}(2)_{MC} \times G_{SM} \ . 
\label{glow}
\eeq
Since the SU(2)$_{HC}$, SU(2)$'_{HC}$, and SU(2)$_{MC}$ gauge interactions
confine, the particles in this effective low-energy field theory are singlets
under all of these groups.  With the fermions having dynamically generated
masses $\ge \Lambda_{MC}$ integrated out, the resultant low-energy effective
SU(2)$_{TC}$ theory consists of the 18 chiral doublets given by
eqs. (\ref{su2doublets}) and (\ref{su2primedoublets}) with the $\chi$'s
removed, namely (with numbers denoting representations under the group of
eq. (\ref{glow})),
\beqs
& & Q_L^t: \ (2;\{1\};3,2)_{1/3,L} \ , \cr\cr
& & U_R^t: \ (2;\{1\};3,1)_{4/3,R} \ , \quad 
    D_R^t: \ (2;\{1\};3,1)_{-2/3,R} \ , \cr\cr
& & L_L^t: \ (2;\{1\};1,2)_{-1,L} \ , \quad 
    E_R^t: \ (2;\{1\};1,1)_{-2,R} \ , \cr\cr
& & \alpha^{1t}_R, \ {\cal N}^t_R, \ \tilde \alpha_{1t,R}: \ 
3(2;\{1\};1,1)_{0,R} 
\label{1fam_fermions_tc} 
\eeqs
where $t=4,5$ refer to SU(2)$_{TC}$ indices and, following standard notation,
we denote technifermions with capital letters. Neglecting the SM interactions,
which are weak at this scale, this theory is vectorial, with nine Dirac
technifermions, including three electroweak-singlet technineutrinos. (Here we
make use of the fact that the representations of SU(2) are (pseudo)real to
re-express the technineutrinos in vectorial form as regards their technicolor
couplings.) The technicolor gauge interaction has the necessary property of
asymptotic freedom, and furthermore, to within the uncertainties inherent in
the analysis of a strong-coupling gauge theory, one may plausibly consider that
it could have walking behavior, so that the technicolor coupling $\alpha_{TC}$,
evolves slowly in the interval below $\Lambda_{MC}$ down to the technicolor
condensation scale.

As the energy scale decreases further to the value that we shall denote 
$\Lambda_{TC}$, the SU(2)$_{TC}$ technicolor theory
produces technifermion condensates that break 
${\rm SU}(2)_L \times {\rm U}(1)_Y$ to U(1)$_{em}$.  These condensates are
\beq
\langle \bar U_{a,t} U^{a,t} \rangle \ , \ 
\langle \bar D_{a,t} D^{a,t} \rangle \ , \ 
\langle \bar E_t E^t \rangle \ , \ 
\label{FFbar}
\eeq
\beq
\langle \bar N_{t,L} \alpha^{1t}_R \rangle + h.c., \ 
\langle \bar N_{t,L} {\cal N}^t_R \rangle + h.c., \ 
\langle \epsilon^{123tv} \bar N_{t,L} \tilde \alpha_{1v,R} \rangle + h.c. 
\label{NNbar}
\eeq
where implied sums are over the color indices $a$ and TC indices $t,v$.  In a
one-family technicolor model such as this, from the relation $m_W^2=(g^2/4)
F_{TC}^2 N_D$ with $N_D=N_c+1=4$ the number of technifermion electroweak
doublets, one has $F_{TC} \simeq 125$ GeV and, taking $\Lambda_{TC} \simeq 2
F_{TC}$, this gives $\Lambda_{TC} \simeq 250$ GeV. The technicolor interaction
also naturally produces Majorana condensates involving only the right-handed,
SM-singlet technineutrinos (which do not break electroweak symmetry), namely,
\beqs
& & \langle \epsilon_{123tv} \alpha^{1t \ T}_R C {\cal N}^{v}_R \rangle + 
h.c., \ \langle \alpha^{1t \ T}_R C \tilde \alpha_{1t,R} \rangle + h.c., \cr\cr
& & \langle {\cal N}^{t \ T}_R C \tilde \alpha_{1t,R} \rangle + h.c.
\label{tcmajorana}
\eeqs
The exact gauge symmetry of the theory at energies below the electroweak scale,
$\Lambda_{TC}$, is
\beqs
& & {\rm SU}(2)_{TC} \times {\rm SU}(2)_{HC} \times {\rm SU}(2)'_{HC} 
\times {\rm SU}(2)_{MC} \times \cr\cr
& & {\rm SU}(3)_c \times {\rm U}(1)_{em} \ . 
\label{exactsymmetry}
\eeqs

The technifermions $F$ involved in the condensates of
eqs. (\ref{FFbar})-(\ref{tcmajorana}) gain dynamical masses that are
generically denoted $\Sigma_F$.  Since other interactions are weaker than
technicolor at this scale, these technifermion condensates are expected to be
nearly equal for different $F$'s, and consequently so are the corresponding
dynamical masses $\Sigma_F$ \cite{lq}. Hence, in particular,
\beq
\frac{|\Sigma_U - \Sigma_D|}{\Sigma_U + \Sigma_D} \ll 1 \ , \quad 
\frac{|\Sigma_N - \Sigma_E|}{\Sigma_N + \Sigma_E} \ll 1 \ . 
\label{sigmadiff}
\eeq
For the overall magnitude of the common $\Sigma_F$, the estimates used, e.g.,
in Ref. \cite{ckm}, give $\Sigma_F \simeq 2 \Lambda_{TC} \simeq 500$ GeV.

Therefore, the model preserves custodial symmetry very well and can naturally
yield acceptably small corrections to the parameter $\rho = \alpha_{em}(m_Z)T$.
An estimate of the contribution in the present model gives a result similar to
the value that we found in Ref. \cite{ckm}.  We recall the reasoning that went
into that estimate.  Let us denote the TC/ETC corrections to $\rho$ as $\Delta
\rho$.  The one-loop ($1\ell$) contribution involving technifermions yields
\beq
(\Delta \rho )_{1\ell} \simeq  \frac{N_{TC} G_F}{8\pi^2 \sqrt{2}} \biggl [ 
N_c f_\rho(\Sigma_U^2,\Sigma_D^2) + f_\rho(\Sigma_N^2,\Sigma_E^2) \biggr ]
\label{deltarho}
\eeq
where \cite{veltman}
\beq
f_\rho(x,y)=x+y-\frac{2xy}{x-y} \, \ln \Big ( \frac{x}{y} \Big ) \ .
\label{fxy}
\eeq
We denote $\Sigma_U-\Sigma_D=\epsilon_{_{UD}}$ and
$\Sigma_N-\Sigma_E=\epsilon_{_{NE}}$.  Since among standard-model gauge
interactions only the weak hypercharge U(1)$_Y$ distinguishes between $U$ and
$D$ (and, separately, $N$ and $E$), one may estimate the standard-model
contribution to these technifermion mass differences as $\epsilon_{_{UD}}
\simeq \epsilon_{_{NE}} \simeq (\alpha_Y/\pi) \Sigma_F$, where $\alpha_Y =
g'^2/(4\pi)$.  With $\alpha_{em}(m_Z) \simeq 1/128$ and $\sin^2\theta_W(m_Z)
\simeq 0.232$, one has $\alpha_Y(m_Z) = 1.0 \times 10^{-2}$, which is also
approximately the value of $\alpha_Y(\mu)$ at a scale $\mu = \Lambda_{TC}$.
This gives a standard-model contribution $\epsilon_{_{UD}}/\Sigma_F \simeq
\epsilon_{_{NE}}/\Sigma_F \simeq 0.3 \times 10^{-2}$.  Other contributions
arise from the different manner in which the ETC and ETC$'$ interactions treat
the $U$ and $D$ (and $N$ and $E$) technifermions (see below). Using the Taylor
series expansion
\beq
f_\rho((\Sigma+\epsilon)^2,\Sigma^2) = \epsilon^2 \bigg [ \frac{4}{3} -
  \frac{1}{15} \frac{\epsilon^2}{\Sigma^2} + 
O \Big (  \frac{\epsilon^3}{\Sigma^3} \Big )  \bigg ] \ , 
\label{fexpand}
\eeq
we can express this contribution as 
\beq
(\Delta \rho )_{1\ell} \simeq 
\frac{N_{TC} G_F}{6 \pi^2 \sqrt{2}}(N_c \epsilon_{_{UD}}^2+\epsilon_{_{NE}}^2) 
\ . 
\label{deltarho_expand}
\eeq
For a rough estimate, setting $\epsilon_{_{UD}} \simeq \epsilon_{_{NE}}$, we
obtain $(\Delta \rho )_{1\ell} \simeq 2 N_{TC}G_F
\epsilon_{_{UD}}^2/(3\pi^2\sqrt{2})$.  Using $N_{TC}=2$ and
$\epsilon_{_{UD}}/\Sigma_{U,D} \simeq 10^{-2}$ then yields $(\Delta \rho
)_{1\ell} \simeq 3 \times 10^{-5}$, or equivalently, $(\Delta T)_{1\ell} \simeq
4 \times 10^{-3}$, which is safely small.  The higher-lying dynamics of
unbroken gauge interactions, such as metacolor, leading to the technicolor
theory should be subsumed in this estimate. Next, one considers contributions
involving explicit ETC and ETC$'$ gauge boson exchanges. These can be roughly
estimated by recalling that the momentum scale of the technicolor mass
generation mechanism is set by $\Lambda_{TC}$, and the emission and
reabsorption of an ETC or ETC$'$ gauge boson will lead to a denominator factor
of at most $1/\Lambda_3^2$. This yields the estimate of Ref. \cite{ckm}, namely
\beq 
\Delta \rho \simeq \frac{2 b \Lambda_{TC}^2}{3 \Lambda_3^2} \ ,
\label{delta_rho} 
\eeq
where $b \simeq O(1)$.  Numerically, this gives $(\Delta \rho) \simeq
(4 \times 10^{-3})b$ or equivalently, $T \simeq 0.5b$.  (We note that this
estimate and the one in Ref. \cite{ckm} are slightly smaller than the one given
in Ref. \cite{met1}.)  For $b \lsim 0.5$, this is consistent with current
experimental constraints \cite{pdg,lepewwg}.  From our studies, this success in
splitting $m_t$ and $m_b$ while plausibly maintaining sufficiently small
corrections to the $\rho$ parameter, appears to generalize beyond just this
particular EMC $L1$ model to other EMC-type ETC models with two ETC gauge
groups.

\subsection{ETC Gauge Bosons}

For a SM fermion $f_\chi$ transforming as a 5 of SU(5)$_{ETC}$, the basic
coupling to the SU(5)$_{ETC}$ gauge bosons (which is vectorial) is
\beq
{\cal L} = g_{_{ETC}} \bar f_j (T_a)^j_k(V_a)^\lambda \gamma_\lambda f^k
\label{gff}
\eeq
where the $T_a$, $a=1,...,24$ are the generators of the Lie algebra of
SU(5)$_{ETC}$ and the $V_a$ are the corresponding ETC gauge fields. For a
fermion $f$ transforming as a 5 of SU(5)$'_{ETC}$, the coupling is the
same with the replacement of $g_{_{ETC}}$ by $g_{_{ETC'}}$ and $(V_a)$ by
$\tilde V_a$.  For nondiagonal transitions, $j \ne k$, it is convenient to use
the fields $V^j_k = \sum_a V_{a,\lambda} (T_a)^j_k$ for SU(5)$_{ETC}$, whose
absorption by $f^k$ yields $f^j$, with coupling
$g_{_{ETC}}/\sqrt{2}$, analogous to the $W^\pm$ in SU(2)$_L$ and similarly for
SU(5)$'_{ETC}$, with the changes noted before.  We take the diagonal (Cartan)
generators for both SU(5)$_{ETC}$ and SU(5)$'_{ETC}$ to be
\beqs
& & T_{24} \equiv T_{d1}=(2\sqrt{10})^{-1}{\rm diag}(-4,1,1,1,1), \cr\cr
& & T_{15} \equiv T_{d2}=(2\sqrt{6})^{-1}{\rm diag}(0,-3,1,1,1), \cr\cr
& & T_8    \equiv T_{d3}=(2\sqrt{3})^{-1}{\rm diag}(0,0,-2,1,1), \cr\cr
& & T_3 = (1/2){\rm diag}(0,0,0,-1,1) \ .  
\label{tds}
\eeqs
The ETC gauge bosons that couple to these diagonal generators $T_{dj}$ are
denoted $V_{dj}$ for SU(5)$_{ETC}$ and $\tilde V_{dj}$ for SU(5)$'_{ETC}$.

\begin{figure}
\begin{center}
\includegraphics[3in, 8in][4in, 9.5in]{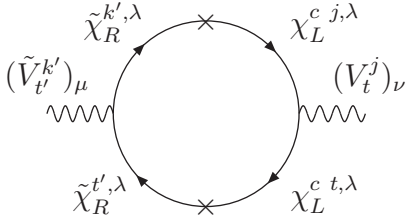}
\end{center}
\caption{\footnotesize{A graph contributing to the ETC gauge boson mixing
$\tilde V^{k'}_{t'} \leftrightarrow V^j_t$, where $j,k' \in \{1,2,3\}$ are
generational indices and $t,t' \in \{4,5\}$ connect with technicolor indices in
SU(2)$_{TC}$, the diagonal subgroup of ${\rm SU}(2)_{ETC} \times {\rm
SU}(2)'_{ETC}$.}}
\label{chigraph}
\end{figure}

When SU(5)$_{ETC}$ breaks to SU(4)$_{ETC}$, the nine ETC gauge bosons in the
coset SU(5)$_{ETC}$/SU(4)$_{ETC}$, namely, $V^1_j$, $(V^1_j)^\dagger=V^j_1$,
$j=2,3,4,5$, and $V_{d1}$, gain masses $M_1 \simeq \Lambda_1$. When
SU(4)$_{ETC}$ breaks to SU(3)$_{ETC}$, the seven ETC gauge bosons $V^2_j$ and
$(V^2_j)^\dagger=V^j_2$, $j=3,4,5$, together with $V_{d2}$, gain masses $\simeq
\Lambda_2$.  Finally, when SU(3)$_{ETC}$ breaks to SU(2)$_{TC}$, the five ETC
gauge bosons $V^3_j$, $(V^3_j)^\dagger=V^j_3$, $j=4,5$, together with $V_{d3}$,
gain masses $\simeq \Lambda_3$.  The analogous statements hold for
SU(5)$'_{ETC}$ with the replacements of $V^j_k$ by $\tilde V^{j'}_{k'}$,
$V_{dj}$ by $\tilde V_{dj'}$, and $\Lambda_j$ by $\Lambda_j'$.  The SM-singlet
fermions responsible for these breakings also, through quantum loops, lead to
mixing among the $V$ bosons and, separately, among the $\tilde V$ bosons, so
that they are not exact mass eigenstates.  The mixing is small, being
suppressed by ratios of the hierarchical ETC and ETC$'$ scales.  There is also
mixing of the ETC and ETC$'$ groups, which takes place at the scale
$\Lambda_{MC}$ via the condensate (\ref{mccondensate}).  This is necessary for
generating down-quark and lepton masses.  A graph that contributes to this
mixing is shown in Fig. \ref{chigraph}.

\subsection{Quark and Lepton Masses}

The effective theory describing the physics at energies $E < \Lambda_{TC}$,
obtained by integrating out the ETC and TC gauge bosons and all of the heavy
fermions, contains the mass matrix of the up-type quarks, 
\beq
{\cal L}_u = -\bar u_{j,L} M^{(u)}_{jk} u^k_R + h.c.,
\label{lmup}
\eeq
and the corresponding mass matrix of the down-type quarks, 
\beq
{\cal L}_d = -\bar d_{j,L} M^{(d)}_{jk'} d^{k'}_R + h.c.,
\label{lmdn}
\eeq
with $j,k' \in \{1,2,3\}$.  For the present model, of L1 type, the mass
matrix for the charged leptons is 
\beq
{\cal L}_e = -\bar e_{j',L} M^{(e)}_{j'k} e^{k}_R + h.c.,
\label{lme}
\eeq
with $j',k \in \{1,2,3\}$. An analogous operator, with obvious interchange of
primed indices, applies for a model of type L2.

\begin{figure}
\begin{center}
\includegraphics[3.5in, 8.7in][4.25in, 9.9in]{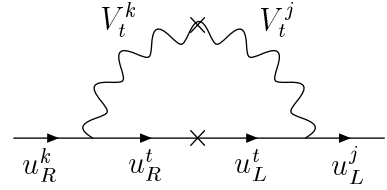}
\end{center}
\caption{\footnotesize{A graph generating $\bar u_{j,L} M^{(u)}_{jk} u^k_R$
where $j,k \in \{1,2,3\}$ are generational indices and $t\in \{4,5\}$ are
technicolor indices. The $V^j_t$ are SU(5)$_{ETC}$ vector bosons.}} 
\label{ugraph}
\end{figure}

The elements of the up-type quark mass matrix arise from the diagram in
Fig. \ref{ugraph}. The diagonal elements of this matrix are 
\beq
M^{(u)}_{jj} \simeq \frac{\kappa \eta \Lambda_{TC}^3}{\Lambda_j^2}
\label{mfu}
\eeq
where $\kappa \sim O(10)$ is a numerical prefactor from the integration (see,
e.g., \cite{ckm}) and $\eta$ is a walking factor
\beq
\eta = \exp[\int_{\Lambda_{TC}}^{\Lambda_w} (d\mu/\mu) \gamma(\alpha(\mu))]
\label{eta}
\eeq
where the technicolor theory has walking behavior between $\Lambda_{TC}$ and a
scale denoted $\Lambda_w$.  With the anomalous dimension $\gamma \simeq 1$ as
in a walking theory, this factor is then $\eta \simeq \Lambda_w/\Lambda_{TC}$.
In the current class of theories, it is plausible that there could be walking
up to the scale $\Lambda_{MC}$, so that $\eta \simeq
\Lambda_{MC}/\Lambda_{TC}$.  For the third-generation standard-model fermions,
the entries $M^{(f)}_{33}$ should give reasonable estimates of the
corresponding masses of $u^3 \equiv t$, $d^3 \equiv b$, and $e^3 \equiv \tau$
which are not changed significantly by off-diagonal entries.  In particular,
for the top quark, assuming the above value of $\eta$ and using the value
$\Lambda_3 \simeq 3$ TeV, one obtains a value of $M^{(u)}_{33}$ and hence $m_t$
that is acceptably close to the experimental pole mass $m_t \simeq 175$
GeV. For $m_t$, the difference between this pole mass and the running mass
evaluated at 175 GeV is negligible; for the other quarks, we use the running
masses evaluated at the scale $\mu_t = 175$ GeV.  In view of the substantial
uncertainties in the dynamically generated standard-model fermion masses due to
the strong-coupling nature of the TC and ETC theories, we consider an estimate
for a fermion mass to be acceptable if it is within a factor of 2-3 of the
measured mass.  If the contributions of off-diagonal element in $M^{(u)}$ are
sufficiently small, then the diagonal element $M^{(f)}_{22}$ is dominant in
determining the mass of the corresponding second-generation fermions $u^2
\equiv c$.  Assuming that this is the case, the value of $\Lambda_2 \simeq 45$
TeV, which is consistent with the renormalization group equations for the
SU(4)$_{ETC}$ theory, yields an acceptable result for $m_c$, i.e., $m_c(\mu_t)
\simeq 0.6$ GeV, corresponding to the pole mass $m_c = 1.3$ GeV. These values
are listed in Table \ref{scales} together with other scales in the model.  The
property $m_u < m_d$ for first-generation quarks, which is the opposite of the
pattern $m_c \gg m_s$, $m_t \gg m_b$ of the second and third generations,
requires that some off-diagonal elements of $M^{(u)}$ and $M^{(d)}$ play an
important role in determining the first-generation quark masses.

\begin{table}
\caption{\footnotesize{Summary table of ETC and ETC$'$ breaking scales and
other strong-coupling scales, in units of TeV, in the model.  In the third
column we list a physical quantity or constraint(s) that are highly correlated
with the numerical value of the given scale. In the lines for $\Lambda_s$ and
$\Lambda_s'$, $f=u,d,e$ refers to up- and down-quark quarks and charged
leptons.  In the line for $\Lambda_{MC}$, the notation $\{m_d\}$, \ $\{m_e\}$
indicates that $\Lambda_{MC}$ affects the values of all of the down-type quarks
and charged leptons.  See text for detailed discussion.}}
\begin{center}
\begin{tabular}{|c|c|c|} \hline\hline
scale & value & comments  \\ \hline
$\Lambda_1$         & $\simeq 10^3$ & FCNC constraints          \\ \hline
$\Lambda_2$         & $\simeq 45$   & $m_c$ value               \\ \hline
$\Lambda_3$         & $\simeq 3$    & $m_t$ value               \\ \hline
$\Lambda_1'$        & $\simeq 10^3$ & FCNC constraints          \\ \hline
$\Lambda_2'$        & $\simeq 15$   & ETC hierarchy             \\ \hline
$\Lambda_3'$        & $\simeq 15$   & $m_b$ value               \\ \hline
$\Lambda_s$         & $\simeq 3$    & $M^{(f)}$, off-diag.      \\ \hline
$\Lambda_s'$        & $\simeq 3$    & $M^{(f)}$, off-diag.      \\ \hline
$\Lambda_{MC}$      & $\simeq 2$    & $\{m_d\}$, \ $\{m_e\}$    \\ \hline
$\Lambda_{TC}$      & $\simeq 0.25$ & $m_W$, \ $m_Z$           \\ \hline\hline
\end{tabular}
\end{center}
\label{scales}
\end{table}

The off-diagonal entries $M^{(u)}_{jk}$, $j \ne k$, arise via diagrams
involving ETC vector boson mixing of the form $V^k_t \to V^j_t$.  This mixing
is indicated by the cross on the ETC vector boson propagator in
Fig. \ref{ugraph}, where it is understood that since the ETC and TC couplings
are strong, further gauge boson exchanges not suppressed by large propagators
are implicitly included.
\beq
M^{(u)}_{jk} \simeq 
\frac{\kappa \eta \ ({}^j_t \Pi^{k}_t) \Lambda_{TC}^3}{\Lambda_j^2 \Lambda_k^2}
\label{mfuoffdiagonal}
\eeq
where ${}^j_t \Pi^k_t$ denotes the relevant nondiagonal ETC propagator
insertion that produce the transition $V^k_t \to V^j_t$.  Additional virtual
SU(5)$_{ETC}$ exchanges not overly suppressed by large mass scales are
understood to be included, since the corresponding gauge couplings are large.

\begin{figure}
\begin{center}
\includegraphics[3.5in, 8.7in][4.25in, 9.9in]{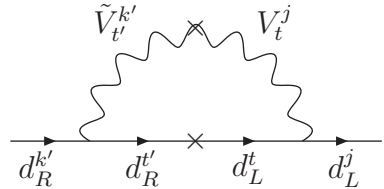}
\end{center}
\caption{\footnotesize{A graph generating the mass matrix for charge $q=-1/3$
quarks, $\bar d_{j,L} M^{(d)}_{jk'} d^{k'}_R$, where $j,k' \in \{1,2,3\}$ are
generational indices and $t,t' \in \{4,5\}$ connect with technicolor indices in
SU(2)$_{TC}$, the diagonal subgroup of ${\rm SU}(2)_{ETC} \times {\rm
SU}(2)'_{ETC}$.  The $V^j_t$ and $\tilde V^{k'}_{t'}$ are SU(5)$_{ETC}$ and
SU(5)$'_{ETC}$ vector bosons; since the corresponding gauge couplings are
strong, this graph is understood to represent also additional exchanges of
these gauge bosons, insofar as they are not suppressed by large mass scales.}}
\label{dgraph}
\end{figure}

The down-quark and charged lepton masses are suppressed, since all elements of
the mass matrices $M^{(d)}_{jk'}$ and $M^{(e)}_{j'k}$ require the $V-\tilde V$
mixing.  The elements of the matrix $M^{(d)}$ arise from the graph in Fig.
\ref{dgraph}.  We have
\beq
M^{(d)}_{jk'} \simeq 
\frac{\kappa \eta \ ({}^j_t \Pi^{k'}_{t'}) \Lambda_{TC}^3}
 {\Lambda_j^2 \Lambda_k'^2}
\label{mfd}
\eeq
where ${}^j_t \Pi^{k'}_{t'}$ denotes the relevant nondiagonal ETC vector bosons
propagator insertions and exchanges that produce the transition $\tilde
V^{k'}_{t'} \to V^j_t$. Again, additional SU(5)$_{ETC}$ and SU(5)$'_{ETC}$
exchanges not overly suppressed by large mass scales are understood to be
included, since the corresponding gauge couplings are large.  Since the
metacolor condensate (\ref{mccondensate}) occurs at the scale $\Lambda_{MC}$,
it follows that the dynamical masses for the metacolor fermions are soft for
higher momentum scales.  Performing the loop integral in Fig. \ref{chigraph},
one thus finds that ${}^j_t \Pi^{j'}_{t'} \propto \Lambda_{MC}^2$ for $j=j'$
(with smaller values for ${}^j_t \Pi^{k'}_{t'}$ for $j \ne k'$). 
The largest eigenvalues of $M^{(u)}$ and $M^{(d)}$, i.e., the masses of $m_t$
and $m_b$, are determined by the respective elements $M^{(u)}_{33}$ and
$M^{(d)}_{33}$.  From eqs. (\ref{mfu}) and (\ref{mfd}), one then has the
relation
\beq
\frac{m_b}{m_t} \simeq \left ( \frac{\Lambda_{MC}}{\Lambda_3'} \right )^2
\label{mbtratio}
\eeq
since the factor of $1/\Lambda_3^2$ cancels in this ratio. 
Using the pole mass $m_b \simeq 4.3$ GeV or equivalently the running mass
$m_b(\mu_t) \simeq 3.0$ GeV, with $m_t \simeq 175$ GeV as above,
eq. (\ref{mbtratio}) yields the ratio $\Lambda_3'/\Lambda_{MC} \simeq
7.6$.  With the illustrative value $\Lambda_{MC} \simeq 2$ TeV, this can be
achieved with $\Lambda_3' \simeq 15$ TeV.  Because of the strong-coupling
nature of the physics involved here, these are only rough estimates, but they
demonstrate that the model can achieve the requisite $t-b$ mass splitting. 

However, the model encounters difficulty in trying to account for the masses of
the standard-model fermions of the lower two generations.  To see this, we
first consider the situation in the approximation where one neglects the
off-diagonal terms in the mass matrices. Then the model would yield the
generalization of eq. (\ref{mbtratio}), viz.,
\beq
\frac{m_{d_j}}{m_{u_j}}=\bigg ( \frac{\Lambda_{MC}}{\Lambda_j'} \bigg )^2
\label{mdjmujratio}
\eeq
for generation $j$.  Taking ratios for $j=2$ and $j=3$, this would imply the
double ratio relation
\beq
\frac{(m_s/m_c)}{(m_b/m_t)}=\left ( \frac{\Lambda_3'}{\Lambda_2'} \right )^2 
\label{doubleratio23}
\eeq
(where all of the masses are evaluated at a common scale, taken here to be
$\mu_t$).  But because $m_s/m_c > m_b/m_t$ (the values being roughly 1/10 and
1/60, respectively), this would require that $\Lambda_3' > \Lambda_2'$, which
is impossible, since the sequential breaking ${\rm SU}(4)'_{ETC} \to {\rm
SU}(3)'_{ETC}$ guarantees that $\tilde \Lambda_2 \ge \tilde \Lambda_3$.  This
problem would be mitigated as much as possible if $\Lambda_2' \simeq
\Lambda_3'$, i.e. the sequential breaking of SU(4)$'_{ETC}$ to SU(3)$'_{ETC}$
and thence to SU(2)$'_{ETC}$ occurs at comparable scales (see Table
\ref{scales}). 

There is also a problem with the ratios of down-quark masses of different
generations.  With the same simplification of neglecting off-diagonal elements
in the mass matrix $M^{(d)}$, one has
\beq
\frac{m_{d_j}}{m_{d_k}} = \bigg ( \frac{\Lambda_k \Lambda_k'}
{\Lambda_j \Lambda_j'} \bigg )^2 \ . 
\label{djdkratio}
\eeq
This yields ratios for $m_s/m_b$ and $m_d/m_s$ that are too small to fit
experimental values.  One is thus motivated to consider the full down-quark
mass matrix in order to assess whether this improves the predictions for the
ratios (\ref{djdkratio}).  We define the ratios
\beq
r_{jk}  \equiv \left ( \frac{\Lambda_j}{\Lambda_k} \right )^2 \ , \quad
r'_{jk} \equiv \left ( \frac{\Lambda_j'}{\Lambda_k'} \right )^2 \ .  
\label{rjk}
\eeq
We can then write eq. (\ref{mfd}) equivalently as 
\beq
M^{(d)} \simeq  m_b \left ( \begin{array}{ccc}
    r_{31}r'_{31}  &  r_{31}r'_{32}   &  r_{31}   \\
    r_{32}r'_{31}  &  r_{32}r'_{32}   &  r_{32}   \\
    r'_{31}        &  r'_{32}         &  1  \end{array} \right ) \ . 
\label{matrixratio}
\eeq
Diagonalizing this matrix, one finds that the presence of the off-diagonal
elements does not change the mass eigenvalues sufficiently from the 
diagonal-matrix case, so that the masses for the first two generations,
$m_d$ and $m_s$, are still too small.  If these off-diagonal elements were
larger than our estimates above, this problem might be ameliorated somewhat.

\begin{figure}
\begin{center}
\includegraphics[3.5in, 8.7in][4.25in, 9.9in]{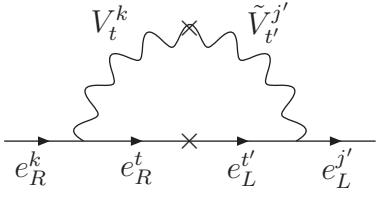}
\end{center}
\caption{\footnotesize{A graph generating the mass matrix for charged leptons,
$\bar e_{j',L} M^{(e)}_{j'k} e^{k}_R$, where $j',k \in \{1,2,3\}$ are
generational indices and $t,t' \in \{4,5\}$ connect with technicolor indices in
SU(2)$_{TC}$, the diagonal subgroup of ${\rm SU}(2)_{ETC} \times {\rm
SU}(2)'_{ETC}$.  Notation is as in Fig. \ref{dgraph}. }}
\label{egraph}
\end{figure}

In this model of type L1, the elements of the charged lepton mass matrix in
eq. (\ref{lme}) are generated via the graph shown in Fig. \ref{egraph}, leading
to the result 
\beq
M^{(e)}_{j'k} \simeq 
\frac{\kappa \eta \, 
({}^{j'}_t \Pi^k_t) \Lambda_{TC}^3}{\Lambda_j'^2 \Lambda_k^2} \ . 
\label{mfe}
\eeq
The model succeeds in producing the desired $t-\tau$ mass splitting, but, as
was the case with $M^{(d)}$ and for the same reasons, the values of $m_e$ and
$m_\mu$ are too small.  For comparison, we note that a similar problem with
overly small $m_{d,s}$ and $m_{e,\mu}$ masses was encountered in the model that
we studied earlier in Refs. \cite{ckm,kt} which used relatively conjugate
representations for the left- and right-handed chiral components of the
down-quarks and charged leptons to obtain intragenerational mass splittings.

In addition to the suppression of the $Q=-1/3$ quark mass relative to the
$Q=2/3$ quark mass in the upper two generations, a viable model should also
incorporate a mechanism for suppressing the charged lepton mass relative to the
$Q=2/3$ and $Q=-1/3$ quark masses in each generation.  In a theory such as the
present one with walking, the technicolor coupling varies slowly with energy
over an extended interval and is only slightly less than the critical value for
condensate formation.  Therefore, small perturbations that would normally be of
negligible importance can become significant.  In particular, the attractive
QCD interaction can naturally expedite techniquark condensation, relative to
technilepton condensation (in a one-family model), so that the former occurs at
a higher energy scale than the latter, giving rise to the inequality $\Sigma_Q
> \Sigma_L$ for the resultant dynamical techniquark and technilepton masses,
and consequently to an increase of the quark masses, relative to the charged
lepton masses, in each generation \cite{qcdcor,met1}.  However, this, by
itself, does not split the masses of the charge 2/3 quarks from those of the
charge $-1/3$ quarks.  For that purpose, one could invoke the U(1)$_Y$
hypercharge gauge interactions, which are weaker.

\section{Some Further Phenomenological Topics}

\subsection{Flavor-Changing Neutral Current Processes}

The present models can satisfy constraints from flavor-changing neutral current
processes.  The basic observation here is the generalization of the point made
in Ref. \cite{ckm} to the case of two ETC groups: because both of the ETC gauge
interactions act in a vectorial, rather than chiral, manner on the
standard-model quarks, processes such as $\bar K^0 \leftrightarrow K^0$, $\bar
D^0 \leftrightarrow D^0$, and $\bar B^0_q \to \bar B^0_q$ with $q=d,s$ are
suppressed.  For example, to take the transition among these that imposes the
most severe constraint, namely the one with neutral kaons, in the $\bar K^0 \to
K^0$ transition, an initial $s_L \bar d_L$ produces a $V^2_1$ ETC gauge boson,
but this cannot directly yield a $d_L \bar s_L$ in the final-state $K^0$; the
latter is produced by a $V^1_2$.  This requires ETC gauge boson mixing of the
form $V^2_1 \to V^1_2$.  In the present models, as in the simpler vectorial
model with a single ETC group studied in Ref. \cite{ckm,kt}, this mixing
introduces a suppression by a factor $\simeq (\Lambda_3/\Lambda_1)^2$,
resulting in an ETC contribution to $\bar K^0 - K^0$ mixing and hence to
$\Delta m_K$ of order $\Lambda_3^2\Lambda_{QCD}^3/\Lambda_1^4$, where $\Delta
m_K = m_{K_L}-m_{K_S} = (0.7 \times 10^{-14})m_{K^0}$, and we ignore factors of
$1/(4\pi^2)$ in view of the strong-coupling nature of the calculation.
Requiring that the ETC contribution be small compared with the experimentally
measured value of $\Delta m_K$ implies that the ratio $\Lambda_3
\Lambda_{QCD}/\Lambda_1^2 \ll 10^{-7}$.  This constraint is satisfied, for
example, by the illustrative values of $\Lambda_1 \simeq 10^3$ TeV and
$\Lambda_3 = 3$ TeV used here, which give $\Lambda_3 \Lambda_{QCD}/\Lambda_1^2
\simeq 0.6 \times 10^{-9}$.  Similarly, in the same $\bar K^0 \to K^0$
transition, an initial $s_R \bar d_R$ produces a $\tilde V^2_1$ ETC gauge
boson, but this cannot directly yield a $d_R \bar s_R$ in the final-state
$K^0$; the latter is produced by a $\tilde V^1_2$.  This requires ETC gauge
boson mixing of the form $\tilde V^2_1 \to \tilde V^1_2$, which causes a
suppression by a factor $\simeq (\Lambda_3'/\Lambda_1')^2$.  By the same
reasoning as above, we require that the ratio $\Lambda_3'
\Lambda_{QCD}/(\Lambda_1')^2 \ll 10^{-7}$.  This constraint is satisfied for
the values of these parameters used here, which give a value of $\simeq 3
\times 10^{-9}$ for this ratio.  Analogous remarks apply for the contributions
via $s_L \bar d_L \to V^2_1 \to \tilde V^1_2 \to d_R \bar s_R$ and $s_R \bar
d_R \to \tilde V^2_1 \to V^1_2 \to d_L \bar s_L$.

The same type of suppression occurs for other neutral pseudoscalar meson
mixings such as $D \leftrightarrow \bar D$, $B_d \leftrightarrow \bar B_d$, and
$B_s \leftrightarrow \bar B_s$.  The upper limit on the decay $K^+ \to \pi^+
\mu^+ e^-$ and hence on the elementary process $\bar s \to \bar d \mu^+ e^-$,
which is mediated by $V^1_2$ and $\tilde V^1_2$, is also satisfied with the
values of $\Lambda_1$ and $\Lambda_1'$ that we use.  Similar consistency checks
can be carried out for other processes and quantities due to dimension-5 and
dimension-6 operators, as we have analyzed these in earlier works
\cite{ckm}-\cite{qdml}.

\subsection{Neutrino Masses} 

In Ref. \cite{nt} a mechanism was presented for producing light neutrino masses
in an ETC theory.  This was studied further in Refs. \cite{lrs,ckm}.  Here we
remark on how this mechanism can be implemented in the current models
containing two ETC gauge groups.  For this purpose, we recall that even in an
ETC theory with only one ETC group, below the highest scale, $\Lambda_1$, of
ETC breaking, there are actually two plausible patterns of breaking.  These
were labelled $G_a$ and $G_b$ in Ref. \cite{nt}, and, in modified form,
sequences S1 and S2 in Refs. \cite{ckm,kt}.  In the discussion above we have
concentrated on the two-ETC group generalization of sequence S1.  For
considerations of neutrino masses, sequence S2 is also of interest.  One would
thus be led to consider a generalization of this sequence to the case of two
ETC groups relevant to the present models.

\subsection{Remarks on a Model with a Minimal Technifermion Sector}

Two continuing concerns that one has with ETC models that contain a full
standard-model family of technifermions are the substantial contribution to the
electroweak parameter $S$ and the presence of many pseudo-Nambu-Goldstone
bosons (PNGB's), whose masses must be elevated (e.g., by walking enhancement)
to lie above experimental lower bounds.  These concerns motivate consideration
of ETC models which use the minimum set of electroweak-nonsinglet
technifermions, comprising (for a given TC index) a single left-handed
SU(2)$_L$ doublet and the corresponding two right-handed fields.  We briefly
comment here on how one might construct a model of this type with two ETC
groups. 

We denote the technicolor contribution to $S$ as $(\Delta S)^{(TC)}$.  In
general, the $S$ parameter \cite{pt} measures heavy-particle contributions to
the $Z$ self-energy via the term $4 s_W^2 c_W^2
\alpha^{-1}_{em}(m_Z)[\Pi^{(NP)}_{ZZ}(m_Z^2) - \Pi^{(NP)}_{ZZ}(0)]/m_Z^2$,
where $s_W^2 = 1-c_W^2 = \sin^2\theta_W$, evaluated at $m_Z$ (see \cite{pdg}
for details).  Since technifermions are strongly interacting on the scale $m_Z$
used in the definition of $S$, one cannot reliably apply perturbation theory to
calculate $(\Delta S)^{(TC)}$ \cite{scalc1,scalc2}; at most, it provides a
rough guide, namely, $(\Delta S)^{(TC)}_{pert.}  = N_D/(6\pi)$, where $N_D$
denotes the total number of new technifermion SU(2)$_L$ doublets (counting
technicolors).  As is well known, for a one-family technicolor with
technifermions transforming according to the fundamental representation of
SU($N_{TC}$), $N_D=N_{TC}(N_c+1)=8$, so that $(\Delta
S)^{(TC)}_{pert.}=4/(3\pi) \simeq 0.4$, which is larger than the experimentally
preferred region.

Now consider a TC/ETC model with a minimal electroweak-nonsinglet technifermion
sector.  For general $N_{TC}$, we again take the technifermions to
transform according to the fundamental representation of SU($N_{TC}$).  The
transformation properties of the SM-nonsinglet technifermions under 
${\rm SU}(N_{TC}) \times G_{SM}$ are given by
\beqs
& & F^{p,t}_L = {F^{1,t}_L \choose F^{2,t}_L} \ : \ 
(N_{TC},1,2)_{0,L}, 
\cr\cr
& & F^{t \ (\pm 1/2)}_R \ : \ \ (N_{TC},1,1)_{\pm 1,R},
\label{1d}
\eeqs
where $p=1,2$ and $t$ are the SU(2)$_L$ and SU($N_{TC}$) indices, the
superscripts in parentheses indicate electric charge, and the subscripts are
hypercharge and chirality.  (The charges on $F^{p,t}_L$ are obvious and hence
are suppressed in the notation.)  For this model, $(\Delta
S)^{(TC)}_{pert.}=N_{TC}/(6\pi)$.  Hence, for $N_{TC}=2$, $(\Delta
S)^{(TC)}_{pert.}=1/(3\pi) \simeq 0.1$, which is sufficiently small to agree
with experimental constraints. Although one-doublet technicolor models, as
such, do not have walking, one can add SM-singlet, TC-nonsinglet fermions so as
to produce walking \cite{ts}.  Another advantage of a one-doublet TC model is
that all of the three Nambu-Goldstone bosons (NGB's) that arise due to the
formation of technicondensates are absorbed to make the $W^\pm$ and $Z$ massive
so that there are no problems with unwanted PNGB's.

In this model, as the energy scale descends to $\Lambda_{TC}$, the TC
interaction naturally leads to the formation of the technifermion condensates
\beq
\langle \bar F_{1,t,L} F^{(1/2),t}_R \rangle \ , \quad
\langle \bar F_{2,t,L} F^{(-1/2),t}_R \rangle
\label{ffbarcond}
\eeq
breaking electroweak symmetry in the desired manner.  This channel has $\Delta
C_2=(N_{TC}^2-1)/N_{TC}$, i.e., 3/2 for $N_{TC}=2$.  We note that for
$N_{TC}=2$ an equally attractive channel would involve the condensate
\beqs
& & \langle \epsilon_{pq} \epsilon_{st} F^{ps \ T}_L C F^{qt}_L \rangle \cr\cr
& = & \langle \epsilon_{st} ( F^{1 s \ T}_L C F^{2 t}_L - 
F^{2 s \ T}_L C F^{1 t}_L )\rangle
\label{ffcond}
\eeqs
where $p,q$ and $s,t$ are SU(2)$_L$ and SU(2)$_{TC}$ indices, respectively.
One would not want (\ref{ffcond}) to be the only technifermion condensate to
occur, since it is invariant under the entire group $G$ in eq. (\ref{g}) and
thus, in particular, does not break the electroweak gauge symmetry.  It would
be worthwhile in future work to investigate how this type of TC model with a
minimal electroweak-nonsinglet technifermion sector could be embedded in a
larger theory with two ETC gauge groups.

\section{Conclusions} 

In this paper we have formulated a class of extended technicolor models that
can plausibly produce the observed mass splitting $m_t \gg m_b$ without
excessive contributions to the $\rho$ parameter or to neutral flavor-changing
processes.  These models use two different chiral ETC groups \cite{met1} such
that left- and right-handed components of charge 2/3 quarks transform under the
same ETC group, while left- and right-handed components of charge $-1/3$ quarks
and charged leptons transform under different ETC groups.  The models suppress
$m_b$ and $m_\tau$ relative to $m_t$, and $m_s$ and $m_\mu$ relative to $m_c$
because the masses of the $Q=-1/3$ quarks and charged leptons require mixing
between the two ETC groups, while the masses of the $Q=2/3$ quarks do not.  We
have constructed and analyzed in detail one explicit model of this type.
Clearly, since the relative sizes of the quark masses in the first generation
are opposite to the order for the higher two generations, i.e., $m_d > m_u$,
the strategy used in these classes of models would not, by itself, be expected
to account for this.  However, it is also difficult to account for the absolute
sizes of $m_s$ and $m_d$, and of $m_\mu$ and $m_e$; the suppression mechanism
makes these smaller than the respective observed values.  A similar problem was
encountered in a model using relatively conjugate representations of a single
ETC gauge group for left- and right-handed chiral components to obtain
intragenerational mass splittings in Refs. \cite{ckm,kt}.  Although the model
is rather complicated, we believe that it is useful as explicit, moderately
ultraviolet-complete realization of the strategy of using two different ETC
groups to account for the splitting between $m_t$ and $m_b$.

\acknowledgments

We thank Prof. T. Appelquist for valuable discussions and Y. Bai for a valuable
comment.  This research was partially supported by the grant NSF-PHY-00-98527.


\begin{thebibliography}{99}
\bibitem{hc} 
The reason that it is necessary to use two different hypercolor groups is
that if one were to try to use just one, it would not be asymptotically free,
and hence the HC interaction would not increase as the energy scale decreases.
Consequently, this HC group would not fulfill its job of helping to produce
various symmetry-breaking bilinear fermion condensates.
\bibitem{efc} An alternative is to avoid a metacolor group and instead have the
communication between the ETC and ETC$'$ groups carried out by one or more
fermions that transform as nonsinglet(s) under both the ETC and ETC$'$ groups;
this will be discussed elsewhere with T. Appelquist, Y. Bai, and M. Piai.
\bibitem{beta}
Our notation for the beta function of a gauge group $G_j$ is indicated below,
where $\alpha_j =g_j^2/(4\pi)$, 
$$
\beta_j = \frac{d \alpha_j}{dt} = - \frac{\alpha_j^2}{2\pi}\left ( b_0^{(j)} +
\frac{b_1^{(j)}}{4\pi}\alpha_j + O(\alpha_j^2) \right )
$$
and $t=\ln \mu$ with $\mu$ being the momentum scale.
\bibitem{gap} 
In the approximation of a single-gauge-boson exchange, the
critical coupling for chiral fermions transforming according to the
representations $R_1$ and $R_2$ of a gauge group $G$ to form a condensate
transforming as $R_{cond.}$ in an SU($N$) gauge theory is given by the
condition $\alpha \ge \alpha_{cr}$, where $\alpha_{cr} \simeq 2\pi/(3\Delta
C_2)$, $\Delta C_2 = [C_2(R_1)+C_2(R_2)-C_2(R_c)]$, and $C_2(R)$ is the 
quadratic Casimir invariant \cite{casimir}.
\bibitem{casimir} 
Our notation for the Casimir invariants $C_2(R)$ and $T(R)$
of the representation $R$ is given by ${\cal D}_R(T_a)^i_j{\cal D}_R(T_a)^j_k =
C_2(R)\delta^i_k$ and $({\cal D}_{\cal R}(T_a))^i_j \ ({\cal D}_{\cal
R}(T_b))^j_i= T({\cal R})\delta_{ab}$, where $\{a,b\}$ and $\{i,j,k\}$ denote
group and representation indices and sums over repeated indices are
understood.
\bibitem{other1} For completeness, we note that there are other condensation
channels with the same value of $\Delta C_2 = 24/5$, such as
$(5,1;1,1,1;3,2)_{1/3,L} \times (\bar 5,1;1,1,1;\bar 3,1)_{-4/3,L} \to
(1,1;1,1,1;1,2)_{-1}$ with condensate $\langle \bar Q_{a,j,L} u^{a,j}_R
\rangle$ and $(5,1;1,1,1;1,1)_{0,R} \times (\bar 5,1;1,1,2;1,1)_{0,R} \to
(1,1;1,1,2;1,1)_0$ with condensate $\langle {\cal N}^{j \ T}_R C
\chi_{j,\lambda,R} \rangle$.  These are both undesired; the former would break
SU(2)$_L$ at too high a scale, and the latter would break SU(2)$_{MC}$.  Our
scenario, in which the condensation (\ref{xixicondensate_model1}) occurs, and
the above two condensations do not occur, at this scale, is consistent with
GMAC arguments, within the uncertainties concerning the properties of strongly
coupled field theories.
\bibitem{lq} 
A possible mechanism for suppressing a common dynamical technilepton mass
$\Sigma_N \simeq \Sigma_E$ relative to a common techniquark mass $\Sigma_U
\simeq \Sigma_D$ is briefly discussed later in the text.  Since this preserves
custodial symmetry, it does not directly contribute to the TC correction to
$\rho$.
\end{thebibliography}
\end{document}